\documentclass[12pt]{article}
\begin{document}
\textwidth = 16 truecm
\textheight = 24 truecm
\hoffset = -2 truecm
\voffset = -2 truecm

\title{\bf{\Large{Study of quantum mechanical scattering in presence of an extra compact space dimension}}}
\author{\bf{\normalsize Ram Narayan Deb}
\thanks{debramnarayan1@gmail.com, debram@rediffmail.com}\\ 
Department of Physics, Chandernagore College,\\
 Chandernagore, Hooghly, West Bengal, Pin-712136, India}
\date{}

\maketitle

\begin{abstract}
It is well known that string theory generates the idea of higher dimensional spacetime instead of the (3+1) dimensions, in which we seem to live. It indicates that the extra space dimensions may remain curled up into very small space. In this paper, we study quantum mechanical scattering in presence of extra compact space dimension, in the hope of devising a method to get a clue to the presence of extra space dimension.  We consider a simplified model of scattering, in which a beam of free particles is scattered due to a one dimensional Dirac-delta potential in presence of an extra compact space dimension. We find that the incident, reflected and transmitted probability current densities of the particles contain the information of the extra dimension. Thus, by measuring experimentally the reflected and transmitted probability current densities of the particles, we may confirm the presence of extra dimension. We also show that the idea of compactification of the extra dimensions gives rise to the quantization of the energy values of the free particles. So, by measuring the quantized energy values of the scattered free particles, we may again confirm the presence of extra dimensions. We also discuss briefly, how the extra compact space dimensions can be detected if the scattering process is performed in (9+1) dimensions, in which six space dimensions are compact.
\end{abstract}

\maketitle
\section{Introduction}
 String theory, which is an excellant candidate for a unified theory of all forces in nature, generates the idea of (9+1) dimensional spacetime instead of (3+1) dimensions \cite{Zwiebach}, \cite{A. Sen}. It suggests that the extra six space dimensions may remain curled up into very small space. The question then arises is, whether it is possible to detect those extra compact dimensions or not, and if possible, then how. In this paper, we study quantum mechanical scattering in presence of extra compact space dimension and try to propose the method of getting a clue to the presence of extra space dimension. 
 We study the quantum mechanical scattering of free particles due to a one dimensional Dirac-delta potential in presence of an extra compact space dimension. We take the delta potential along $z$-direction and take the $y$-dimension to be curled up into a small circle of radius $R$. The resulting space is a cylinder of radius $R$ and of infinite length along the $z$-direction. We consider a beam of free particles moving along the positive $z$-axis and incident on the delta potential at $z=0$. We find that the incident, reflected and transmitted probability currents of the particles depend upon the extra compact dimension. Thus, by measuring experimentally the reflected and transmitted probability currents of the particles, we may get a clue to the presence of extra dimension. We also 
show that, the idea of compactification of the extra dimension leads to the quantized energy values of the free
particles and thus, by measuring the quantized energy 
values of the scattered particles, we may confirm the 
presence of extra dimension.

 In section 2 we present our study on quantum mechanical scattering of free particles in one dimension in presence of an extra compact space dimension, in section 3 we predict the results of scattering experiments done in (9+1) dimensions in which six space dimensions are compact, and in section 4 we present the summary and conclusion.

\section{Quantum mechanical scattering due to a one dimensional Dirac-delta potential in presence of an extra compact dimension} 

We consider a Dirac-delta potential $\lambda\delta(z)$, of strength $\lambda$, along the $z$-axis in presence of an extra compact dimension. The extra dimension that we take here is along the $y$-direction. We assume that the $y$-direction is curled up into a small circle of radius $R$. Therefore, the resulting space is a cylinder of radius $R$ and infinite length along the $z$-direction.  The delta potential divides the $z$-axis into two 
regions, one on the left side of the delta potential, which is named as region I and the other on the right side of the delta potential, which is named as region II. 

We consider a beam of free particles each of mass $m$ and energy 
$E$ moving along positive $z$-axis in region I and incident on the delta potential at $z=0$. If the wave function of a particle in region I is $\psi_I(y,z)$, the time independent 
Schr$\ddot{o}$dinger equation for the particle in region I is
\begin{equation}
-\frac{\hbar^2}{2m}\nabla^2\psi_I(y,z) = E\psi_I(y,z).
\label{26}
\end{equation}  
Since the particles here are moving on the surface of a cylinder, it is convenient to use cylindrical coordinates to solve 
Eq. (\ref{26}). The rectangular cartesian coordinates $(x, y, z)$ are related to the cylindrical coordinates $(\rho, \phi, z)$ as
$x = \rho\cos\phi$, $y = \rho\sin\phi$ and $z=z$.

Since the particles are confined on the surface of a cylinder of radius $R$ here, hence, Eq. (\ref{26}) in cylindrical coordinates reduces to
\begin{equation}
-\frac{\hbar^2}{2m}\bigg{[}\frac{1}{R^2}\frac{\partial^2\psi_I(\phi,z)}{\partial\phi^2} + \frac{\partial^2\psi_I(\phi,z)}{\partial z^2} \bigg{]} = E \psi_I(\phi,z).
\label{27}
\end{equation}

We use the method of separation of variables, and take
\begin{equation}
\psi_I(\phi,z) = \Phi_I(\phi)Z_I(z).
\label{28}
\end{equation}
Using this in Eq. (\ref{27}), we get 
\begin{equation}
\frac{\hbar^2}{2mR^2}\frac{1}{\Phi_I(\phi)}
\frac{d^2\Phi_I(\phi)}{d\phi^2} + E = -\frac{\hbar^2}{2m}\frac{1}{Z_I(z)}\frac{d^2Z_I(z)}{dz^2}.
\label{29}
\end{equation} 
Both sides of the above equation must be a constant and let the constant be $E_1$.
So, we get
\begin{equation}
\frac{d^2Z_I(z)}{dz^2} = -\frac{2mE_1}{\hbar^2}Z_I(z).
\label{30}
\end{equation}
Let
\begin{equation}
\frac{2mE_1}{\hbar^2} = k_1^2.
\label{31}
\end{equation}
Using this in Eq. (\ref{30}), we get the solution of the equation as
\begin{equation}
Z_I(z) = A_1e^{ik_1z} + B_1e^{-ik_1z}.
\label{32}
\end{equation}
Similarly, solving the time independent Schr$\ddot{o}$dinger equation for the particle in region II, we obtain
\begin{equation}
Z_{II}(z) = C_1e^{ik_1z} + D_1e^{-ik_1z}.
\label{33}
\end{equation}
Since, there is no leftward moving particle in region II, the constant $D_1=0$.
Therefore,
\begin{equation}
Z_{II}(z) = C_1e^{ik_1z}.
\label{34}
\end{equation}
The $\phi$-part of the Schr$\ddot{o}$dinger equation in region I is
\begin{equation}
\frac{\hbar^2}{2mR^2}\frac{1}{\Phi_I(\phi)}\frac{d^2\Phi_I(\phi)}
{d\phi^2} + E = E_1.
\label{35}
\end{equation} 
Using
\begin{equation}
\frac{2m}{\hbar^2}R^2(E-E_1) = k_2^2,
\label{36}
\end{equation}
we obtain the solution of Eq. (\ref{35}) as
\begin{equation}
\Phi_I(\phi) = F_1\cos k_2\phi + G_1\sin k_2\phi,
\label{37}
\end{equation}
where $F_1$ and $G_1$ are constants.
Since, $\phi$ and $\phi + 2\pi$ are the same points we must have
\begin{equation}
\Phi_I(\phi + 2\pi) = \Phi_I(\phi).
\label{38}
\end{equation}
Therefore,
\begin{equation}
F_1\cos k_2(\phi + 2\pi) + G_1\sin k_2(\phi + 2\pi) = F_1\cos k_2\phi + G_1\sin k_2\phi.
\label{39}
\end{equation}
This condition yields
\begin{equation}
k_2 = n,
\label{39a}
\end{equation}
where $n$ is an integer. Therefore,
\begin{equation}
\Phi_I(\phi) = F_1\cos n\phi + G_1\sin n\phi.
\label{40}
\end{equation}
Therefore, the total wave function of the particle in region I is
\begin{equation}
\psi_{I}(\phi, z) = (A_1e^{ik_1z} + B_1e^{-ik_1z})( F_1\cos n\phi + G_1\sin n\phi).
\label{40a}
\end{equation}
Now, since the delta potential has no dependence on $\phi$ (in cartesian coordinates it is independent of $y$), the 
$\phi$-part of the wave function in region II should be same as that in region I.
So,
\begin{equation}
\Phi_{II}(\phi) = \Phi_I(\phi).
\label{41}
\end{equation}
Therefore, the total wave function of the particle in region II is
\begin{equation}
\psi_{II}(\phi,z) = C_1e^{ik_1z}(F_1\cos n\phi + G_1\sin n\phi).
\label{54} 
\end{equation} 
Now the boundary conditions on the wave functions are
\begin{equation}
\psi_{I}(\phi,z=0) = \psi_{II}(\phi, z=0)
\label{42}
\end{equation}
and
\begin{equation}
\frac{d\psi_{II}}{dz}\Bigg{\vert}_{z=0} - \frac{d\psi_{I}}{dz}\Bigg{\vert}_{z=0} = \frac{2m}{\hbar^2}\lambda\psi_{II}(\phi,0).
\label{43}
\end{equation}

 Using the boundary condition, that is, Eq. (\ref{42}) on 
$\psi_{I}(\phi,z)$
and $\psi_{II}(\phi,z)$, we obtain
\begin{equation}
A_1 + B_1 = C_1.
\label{55}
\end{equation}
The condition of discontinuity of the first derivative of the wave functions at $z=0$, that is, Eq. (\ref{43}) produces,
\begin{equation}
ik_1C_1 - ik_1(A_1 - B_1) = \frac{2m}{\hbar^2}\lambda C_1.
\label{56}
\end{equation}

Solving Eqs. (\ref{55}) and (\ref{56}), we obtain
\begin{equation}
B_1 = -i\frac{(m/\hbar^2)\lambda}{k_1+i(m/\hbar^2)\lambda}A_1
\label{57}
\end{equation}
and
\begin{equation}
C_1 = \frac{k_1}{k_1+i(m/\hbar^2)\lambda}A_1.
\label{58}
\end{equation}

We now proceed to calculate the incident, reflected and transmitted probability current densities.These are respectively,
\begin{eqnarray}
\overrightarrow{J_i} &=&\frac{i\hbar}{2m}\Big{[}\psi_i\overrightarrow{\nabla}{\psi_i}^\star - {\psi_i}^\star\overrightarrow{\nabla}\psi_i\Big{]},\label{59}\\
\overrightarrow{J_r} &=&\frac{i\hbar}{2m}\Big{[}\psi_r\overrightarrow{\nabla}{\psi_r}^\star - {\psi_r}^\star\overrightarrow{\nabla}\psi_r\Big{]},\label{60}\\
\overrightarrow{J_t} &=&\frac{i\hbar}{2m}\Big{[}\psi_t\overrightarrow{\nabla}{\psi_t}^\star - {\psi_t}^\star\overrightarrow{\nabla}\psi_t\Big{]}.\label{61}
\end{eqnarray}
From Eqs. (\ref{40a}) and (\ref{54}), we see that
\begin{eqnarray}
\psi_i &=& A_1e^{ik_1z}(F_1\cos n\phi + G_1\sin n\phi),\label{62}\\
\psi_r &=& B_1e^{-ik_1z}(F_1\cos n\phi + G_1\sin n\phi),\label{63}
\\
\psi_t &=& C_1e^{ik_1z}(F_1\cos n\phi + G_1\sin n\phi).\label{64}
\end{eqnarray}
Using Eqs. (\ref{59}) and (\ref{62}), we obtain
\begin{eqnarray}
\overrightarrow{J_i} &=& \frac{\hbar}{2m}|A_1|^2\Big{[}
\hat{\phi} \frac{in}{R}(F_1G_1^\star - F_1^\star G_1) + \hat{z}2k_1(|F_1|^2\cos^2n\phi\nonumber\\
&+& \frac{1}{2}(F_1G_1^\star + F_1^\star G_1)\sin 2n\phi + |G_1|^2\sin^2n\phi) \Big{]}.
\label{65}
\end{eqnarray}
Similarly, using Eqs. (\ref{60}) and (\ref{63}), we get
\begin{eqnarray}
\overrightarrow{J_r} &=& \frac{\hbar}{2m}|B_1|^2\Big{[}
\hat{\phi} \frac{in}{R}(F_1G_1^\star - F_1^\star G_1) - \hat{z}2k_1(|F_1|^2\cos^2n\phi\nonumber\\
&+& \frac{1}{2}(F_1G_1^\star + F_1^\star G_1)\sin 2n\phi + |G_1|^2\sin^2n\phi) \Big{]}.
\label{66}
\end{eqnarray}
In the same fashion, using Eqs. (\ref{61}) and (\ref{64}), we get
\begin{eqnarray}
\overrightarrow{J_t} &=& \frac{\hbar}{2m}|C_1|^2\Big{[}
\hat{\phi} \frac{in}{R}(F_1G_1^\star - F_1^\star G_1) + \hat{z}2k_1(|F_1|^2\cos^2n\phi\nonumber\\
&+& \frac{1}{2}(F_1G_1^\star + F_1^\star G_1)\sin 2n\phi + |G_1|^2\sin^2n\phi) \Big{]}.
\label{67}
\end{eqnarray}
Thus, we observe that the incident, reflected and transmitted 
probability current densities depend on $\phi$, that is on extra dimension. We also observe that $\overrightarrow{J_i}$, 
$\overrightarrow{J_r}$ and $\overrightarrow{J_t}$ are real quantities.
Now, using Eqs. (\ref{65}), (\ref{66}) and (\ref{67}), we obtain the reflection and transmission coefficients as
\begin{equation}
R_1 = \frac{|\overrightarrow{J_r}|}{|\overrightarrow{J_i}|} = \frac{|B_1|^2}{|A_1|^2} = \frac{(m^2/\hbar^4)\lambda^2}{k_1^2 + 
(m^2/\hbar^4)\lambda^2}
\label{68}
\end{equation}
and
\begin{equation}
T_1 = \frac{|\overrightarrow{J_t}|}{|\overrightarrow{J_i}|} = \frac{|C_1|^2}{|A_1|^2} = \frac{k_1^2}{k_1^2 + 
(m^2/\hbar^4)\lambda^2}
\label{69}
\end{equation}
respectively.
We observe that $R_1 + T_1 = 1$, as it should be, proving the conservation of the total number of particles.

We now concentrate on Eqs. (\ref{36}) and (\ref{39a}). Combining the two equations, we get
\begin{equation}
E = E_1 + \frac{n^2\hbar^2}{2mR^2}.
\label{70}
\end{equation} 
Using Eqs. (\ref{31}) and (\ref{70}), we obtain
\begin{equation}
E = \frac{\hbar^2k_1^2}{2m} + \frac{n^2\hbar^2}{2mR^2}.
\label{71}
\end{equation} 
We see from the above equation that the first term on the right hand side is like the energy of a particle in absence of any extra compact dimension. Due to the presence of the second term on the right hand side of the above equation, we see that the energy values of the free particles are modified and this modification is due to the presence of the extra compact dimension. Now, we see that since, $n$ is
an integer, the second term on the right hand side has quantized values. Thus, due to the presence of an extra compact dimension, the energy values of a free particle are quantized.  The origin of this quantization is the periodicity condition, that is Eq. (\ref{38}), which must be satisfied by the 
$\phi$-part of the wave function of the particle. We see that if $R$ is very small the second term on the right hand side of the above equation has very large value. Thus, if the extra dimension is compactified into a very small space, the modified part of the energy values of the free particle lies on a very high energy scale. The interesting fact is that the energy values of the free particles are quantized due to the presence of the  extra compact space dimension.

After gathering experience by analysing the quantum mechanical scattering process in (2+1) dimensions, where one of the space dimensions is compactified, we now proceed to guess the possible outcome of the scattering process performed in (9+1) dimensional spacetime, in which  six space dimensions are compact.

\section{Prediction of the results of scattering experiment performed in (9+1) dimensions, in which six  space dimensions are compact}

From the above analysis of the scattering process performed in (2+1) dimensional spacetime, where one of the space dimensions is compact, we may guess some important results, regarding the scattering process performed in (9+1) dimensional spacetime, in which six space dimensions are compact.

We may guess that in such scattering process, the probability current densities of the scattered particles
will depend upon the coordinates along the six extra compact space dimensions. Thus, by measuring the probability current densities of the scattered particles, we may confirm the presence of extra dimensions.

We have also seen from the calculations performed in the previous section, that the energy values of the scattered particles are quantized in presence of an extra compact dimension. From Eq. (\ref{71}), we see that the single extra compact space dimension adds the extra term 
$\frac{n^2\hbar^2}{2mR^2}$, which has quantized values. Therefore, we can guess that if scattering experiment is performed in (9+1) dimensions, in which six space dimensions are compact we will obtain six more terms, added to the energy value of the particle in (3+1) dimensions, corresponding to the six extra compact dimensions. These six extra terms will have quantized values, because the wave function of a particle must satisfy the periodicity conditions like Eq. (\ref{38}), along the respective extra compact diemnsions. All these extra terms will become prominent in the ultra-high energy scale. Therefore, in the high energy scattering experiments, if we observe quantized energy values of the scattered particles, we may confirm that this is due to the presence of extra compact space dimensions. This is another way of detecting the presence of extra compact space dimensions.

\section{Summary and conclusion}

We investigated the quantum mechanical scattering of free particles due to a one dimensional Dirac-delta potential in presence of an extra compact dimension. We found that the incident, reflected and transmitted probability current densities of the particles depend on the extra dimension. Thus, by measuring experimentally the reflected and transmitted probability currents of the particles, we may get a clue to the presence of extra dimension.
We have also discussed that in (9+1) dimensions, in which six space dimensions are compact, the probability current densities of the scattered particles will depend on the coordinates along the six extra dimensions. Thus, by analysing the probability current densities of the scattered particles in (9+1) diemnsions, we may confirm the presence of extra space dimensions.   

We have also seen that the presence of a single extra compact space dimension gives rise to the quantized energy values of the scattered particles. Thus, by measuring the quantized energy values of the scattered particles we may confirm the presence of the extra space dimension.
 We have also discussed that, in case of scattering process performed in (9+1) dimensions, in which six space dimensions are compact the energy values of the particles will get modified due to the addition of six extra quantized terms. So, by observing the quantized energy values of the scattered particles in (9+1) dimensions, we may confirm the presence of extra compact space dimensions.

 We hope that our study on this simple model of scattering in presence of extra compact dimensions
may bring some new insights into the method of experimental verification of extra compact space dimensions, which are predicted by string theory.

{\vskip 1cm}
{\Large\bf Acknowledgements}
{\vskip 0.5cm}

I am grateful to Kuntal Gupta and Bishwajit Paul for helping me to install the relevant softwares in my personal computer for the preparation of this manuscript.


\begin{thebibliography}{101}
\bibitem{Zwiebach} B. Zwiebach, A First Course in String Theory, Second Edition, Cambridge University Press, 2009.
\bibitem{A. Sen} A. Sen, Recent Developments in Superstring Theory, arXiv:hep-lat/0011073v2.
   
\end{thebibliography}
\end{document}